# The melting curve of iron at extreme pressures: implications for planetary cores

G.Morard[1,2], J.Bouchet[3], D.Valencia[4], S.Mazevet[3], F.Guyot[1,2]

[1] *Institut de Minéralogie et de Physique des Milieux Condensés (UMR 75-90), 140 rue de Lourmel, 75015 Paris, France*

[2]*Institut de Physique du Globe de Paris, 4 Place Jussieu, 75252 Paris, France*

[3]*CEA, DAM, 91297 Arpajon, France*

[4] *Observatoire de la Cote d'Azur, 06560 Valbonne, France*

**Abstract**

Exoplanets with masses similar to that of Earth have recently been discovered in extrasolar systems. A first order question for understanding their dynamics is to know whether they possess Earth like liquid metallic cores. However, the iron melting curve is unknown at conditions corresponding to planets of several times the Earth's mass (over 1500 GPa for planets with 10 times the Earth's mass ($M_E$)). In the density-temperature region of the cores of those super-Earths, we calculate the iron melting curve using first principle molecular dynamics simulations based on density functional theory. By comparing this melting curve with the calculated thermal structure of Super Earths, we show that planets heavier than $2M_E$, have solid cores, thus precluding the existence of an internal metallic-core driven magnetic field. The iron melting curve obtained in this study exhibits a steeper slope than any calculated planetary adiabatic temperature profile rendering the presence of molten metallic cores less likely as sizes of terrestrial planets increase.

**Introduction**

Several exoplanets with masses similar to that of Earth have been detected (Beaulieu et al., 2006) and studies have speculated (Valencia et al., 2006) about the state of the iron cores of extrasolar terrestrial planets larger than Earth, the so called super-Earths. In order to better constrain the internal dynamics of these planets, accurate knowledge of the high pressure melting curve of iron is required. Current models suggest that internal pressures in these planets can extend over 1500 GPa (Valencia et al., 2006). Properties of pure iron, which is the main component of the planetary cores, as well as of alloying with light elements which tend to decrease its melting point (Poirier, 1994), are thus needed under such extreme conditions.

The high pressure melting curve of iron at conditions ranging from a few GPa's to pressures relevant to the Earth center (360 GPa) has previously been studied using shock waves, e.g. (Brown and McQueen, 1986; Nguyen and Holmes, 2004; Yoo et al., 1993), and laser-heated diamond anvil cells, e.g. (Boehler, 1993; Shen et al., 1998). The significant uncertainties of the experimental results obtained at such extreme pressures have generated an intense activity on the theoretical side. Several predictions of the high pressure melting curve of iron have been published over the past ten years using either classical potentials or fully quantum description of the electrons based on density functional theory. Concerning terrestrial exoplanets, the internal conditions predicted, of over 1500 GPa for planets with 10 $M_E$ (Valencia et al., 2006), together with temperatures of a few thousands Kelvin, are intractable by existing experimental techniques: static compression methods do not reach such high pressures whereas shockwaves bring temperatures that are far too high at the relevant pressures. In spite of being computationally extremely expensive, *ab initio* molecular dynamics (AIMD) simulations are currently the best practical mean to obtain useful information on the iron properties under such pressure-temperature conditions. Using classical potentials, Belonoshko et al. (Belonoshko et al., 2000) predict a melting point of 7100 K at the pressure of the Earth's Inner Core Boundary (ICB). With thermodynamic integrations, Alfè et al. (Alfè et al., 2002b) obtain a temperature of 6400 K, in agreement with calculations using solid-liquid coexistence by AIMD (Alfè, 2009) (Figure 1). Recently, using Monte Carlo free energy calculations, Sola and Alfè (2009) obtain a melting temperature of 6900 K at the ICB.

**Computational methodology**

Simulations were performed using the **ABINIT** package (Gonze et al., 2002) where the electrons are treated quantum mechanically using density functional theory (DFT) while the ions are advanced classically using the resulting forces. The calculations used the projected augmented wave (PAW) (Blochl, 1994; Torrent et al., 2008) method for the calculation of the electronic structure and the generalized gradient approximation (GGA) according to the recipe of Perdew, Burke and Ernzerhof for the exchange-correlation energy and potential (Perdew et al., 1996). We generated a pseudo-potential with 3$s$, 3$p$, 3$d$, and 4$s$ states as valence electrons (see (Dewaele et al., 2008) for a detailed description of the generation of the pseudopotential and a comparison with recent experimental data). The simulations were performed using a cutoff energy, $E_c$, for the plane wave basis chosen equal to 350 eV. The radius of the augmentation regions for the PAW pseudo-potential was chosen small enough to avoid an overlapping of the spheres surrounding each atom at the highest densities. An efficient scheme for the parallelization was used to perform the simulations involving a large number of atoms and time steps (Bottin et al., 2008). Simulations presented here were carried out in the *NVT* ensemble, where *N, V, T* stand for, respectively, the number of particles, volume and temperature. We systematically used an electronic temperature equal to the ionic temperature in our simulations.

The calculations of the melting points (Figure 1) were performed at four different densities, 13.15, 16, 18 and 20 g/cm$^3$ to cover the pressure range expected in super-Earths. Two different simulation techniques can be used to obtain the melting curve of a material using AIMD. The so called "heat until it melts" (HUM) approach which is a one phase approach, and the two-phases and coexistence methods (See (Bouchet et al., 2009)). In the HUM method, a supercell representing the solid phase expected close to the melting line is heated gradually until melting occurs. In the two-phases method, simulations are performed starting with a supercell containing two phases in equal amounts, one solid and one liquid. Note that contrary to the phase coexistence approach where simulations are performed in the NVE ensemble and where the solid and liquid states can coexist (Alfè, 2009), our two-phases simulations (e.g. (Belonoshko, 1994)), performed in the NVT ensemble, tend to a single phase: above Tm, the system will become liquid, whereas below Tm, the liquid part will solidify. In the HUM method, which is less expensive computationally as it requires smaller simulation times and smaller supercells, the crystal is heated homogeneously which leads to a significant over-estimation of the melting temperature. In contrast, the melting is heterogeneous in the two-phases method due to the solid-liquid interface and thus no overheating occurs. The two methods can result in substantially different melting temperatures. The two-phases method generally yields a better estimate of the melting temperature, Tm, obtained by equating the Gibbs free energies, while the HUM approach can overestimate Tm by as much as 30% depending on the pressure and on the compound studied (Bouchet et al., 2009). Despite this drawback, the computational efficiency of the HUM method makes it attractive to obtain an initial estimate of the melting curve (Raty et al., 2007). In our calculations, we first used the HUM method to test the effect of the structure of iron and to determine which phase should be used in the more computationally expensive two-phases method, which was then applied.

We performed simulations for both the hcp and bcc structures since these two phases are candidates for the high pressure stable phase of iron (Alfè et al., 2001; Belonoshko et al., 2003). Our results are presented in Fig. 2 for supercells using 54 (3×3×3) atoms and four k-points and 128 (4×4×4) atoms at the Γ point. These samplings ensure that the uncertainties in the pressure were within 1%. First, we observed a large discrepancy in the prediction of the melting curve between the bcc and hcp phases when using only 54 atoms and four k-points in the simulations cell. These simulation parameters correspond to the ones previously used to calculate the elastic constants with the AIMD method (Vocadlo et al., 2003). For simulations using 128 atoms, which have similar cell sizes as the ones used for calculations performed for other materials such as Mo (Belonoshko et al., 2008) or Na (Raty et al., 2007), we now find similar melting curves when using either the bcc or the hcp structures as initial solid phase (Figure 2). The simulations using the bcc phase show a slightly larger melting temperature in the low pressure range but, at high pressures, above 1000 GPa, the two melting curves tend to the same values. This is in agreement with previous calculations using classical potentials fitted on *ab initio* results (Belonoshko et al., 2003). As the addition of light elements seems to favor the formation of a bcc phase (Belonoshko et al., 2009; Cote et al., 2008; Dubrovinsky et al., 2007), and due to the little difference found in the HUM method, we decided to use the bcc structure in our two-phases calculations to obtain a precise melting curve of iron at high pressure. The difference induced by this phase difference is small in regard of the uncertainties introduced by the addition of alloying elements in iron.

**Results**

Our results using these approaches are presented in Fig 1 and compared to previous calculations performed at conditions corresponding to the Earth's inner core boundary (ICB). Alfè *et al.* (Alfè et al., 2002b) used the free energy approach (FEA) while Belonoshko *et al.* (Belonoshko et al., 2000) used the interface method with an embedded-atom method (EAM) potential fitted to *ab initio* calculations. Although differences exist between the results obtained by these two groups, Alfè *et al.* (Alfè et al., 2002a) have shown how to reconcile the EAM results to recover the melting curve based on free energy. Since our calculations are fully *ab initio,* they do not suffer from this problem. We have used the two-phases method with cells containing 108 (2x54) and 256 (2x128) atoms, see Fig 2. As observed in aluminum (Bouchet, 2009), the discrepancies between the two cells increase with pressure, around 7% for the highest pressure considered here, while for the lowest pressure the difference is between the error bars. At the terrestrial ICB conditions, see inset of Fig 1, the melting point is in excellent agreement with the most recent calculations obtained using the coexistence approach and a larger super-cell of 1000 atoms (Alfè, 2009). This gives us strong confidence in the reliability of the melting curve calculated using this approach with 256 atoms in the simulation cell, in spite of the resulting larger error bar on the melting temperature calculated. In the current work, we complement these calculations with simulations extending to the expected pressures in super Earths of 10 times the mass of our planet (1500 GPa). As expected, we observe a strong reduction of the melting temperature between the HUM and the two-phases approach. Also, we do not find significant differences between melting curves calculated using either a bcc or a hcp structure in the HUM method, up to the highest pressures explored here. The two-phases approach leads to melting temperatures almost constantly increasing as a function of pressure. The slope is, however, slightly smaller than when calculated with the HUM method and yielding a melting temperature at 1500 GPa of about twice the one obtained at terrestrial ICB conditions. This contrasts with findings in systems such as Na and H where maxima in the melting curve were identified at high pressures (Bonev et al., 2005; Raty et al., 2007). The melting curve behavior found here has direct implications for the state of iron in super-Earth' cores.

**Thermal structure of exoplanets and implications for the Super-Earths' cores**

The thermal structures of the Earth and Earth-like planets are based on nearly adiabatic average temperature profiles (Brown and Shankland, 1981), an assumption likely to be valid in convective systems (Tozer, 1972). The adiabatic profile is anchored below a conductive lid, the lithosphere, on the shallower side, at a temperature sufficient to allow enough plasticity of the rocks, which mainly depends on the rheology of mantle materials. Models for the thermal structure of the Earth are usually based on geological evidences of temperatures of ~1300 K at about 100 km depth (Valencia et al., 2006). The usual assumption is that this should not be too different in other telluric planets since the chemical compositions are believed to be close. In the Earth, a quasi adiabatic gradient anchored at ~ 1300 K is consistent with the confrontation of laboratory data with seismological evidence that the ringwoodite-silicate perovskite transition occurs a temperature of ~1873 K at 670 km (~25 GPa) (Ito and Katsura, 1989). Once an adiabatic profile is anchored at T=1873 K-670 km, temperatures of 2773 K (Brown and Shankland, 1981) are calculated at the Core-Mantle Boundary (CMB) of the Earth, on the mantle side. A model from Valencia (Valencia et al., 2007) gives temperatures slightly lower (around 2500 K) but consistent : unknown values of the Grüneisen parameters, in materials such as silicate post-perovskite or low spin bearing iron minerals (Badro et al., 2004) may affect these estimates by uncertainties of few hundreds of K, which are usually considered as not larger than possible deviations from isentropy in convective media. The last major unknown is related to the thermal boundary layers, either internal to the mantle or, more important, at the planetary CMB. The latter one is closely related to the production of heat within the core. Indeed, the temperature jump at the CMB is not well constrained in the Earth itself. Possible variations between 1350 K and 350 K have been proposed (Leitch, 1995), leading to temperatures of up to 3700 K in the Earth's core at the CMB, for the most recent model based on experimental constraints (Tateno et al., 2009). Recent geotherm calculations, based on the double-crossing of post-perovskite phase transition (Hernlund and Labrosse, 2007), show a large temperature jump of ~2500 K, in order to match with iron melting curve from (Alfè et al., 2002b).

In Figure 3, the thermal profiles from (Valencia et al., 2007) are given for a planet of 1 $M_E$. This model has been modified, considering that, adding all possible thermal boundary layers, core temperatures should not exceed the planetary isentrope by more than 1500 K as materialized in Figure 3 by a temperature jump of 1500 K at the core mantle boundary. In order to be conservative, an additional error bar of 1000 K (+/- 500 K) was added as an attempt to take into account the aforementioned uncertainties of anchoring the isentrope temperatures, Grüneisen parameters and deviations to isentropy. This leads to possible temperature profiles in Earth and in super-Earths with 5 and 10 Earth masses as given in Figure 3.

However, considering exoplanets could lead to several qualitative differences with the thermal profile of the Earth. First, the age of the planet controls its thermal state. However, if the same cooling rate as the Earth is applied for exoplanets (70 K/Gyr (Davies, 2007)), the time dependence of the exoplanets internal temperature would stay within the estimated error bars. Secondly, depending on the orbital distance and irradiation from the star, temperature at the surface could be largely variable. For example, the surface temperature for CoRoT-7b, a super-Earth recently discovered, is estimated to be around 2000 K (Leger et al., 2009). Such hot temperature profiles are considered in (van den Berg et al., 2010), as shown in Figure 3. Concerning this model, the highest temperature profile, with 10 000 K at the CMB for a planet with 8 $M_E$, seems unlikely, as partial melting should thus occur in the mantle, by comparison with the solidus of silicates (Stixrude et al., 2009). A partially molten mantle would likely be associated with rapid planetary cooling, thus leading to silicate solidification. Third, the concentration of radioactive heat sources may be different than for Earth, especially in terms of potassium given its volatile character. Four, the interior of terrestrial super-Earths, and in particular the mantle, is only known to the extent of the Earth's interior which only reaches the 136 GPa pressure at the CMB. This means that any possible additional phase transition in the mantle at higher pressures than the perovskite/post-perovksite phase change, in particular the controversial $MgSiO_3$ dissociation (Grocholski et al., 2010; Umemoto et al., 2006), could occur in the deep mantle of exoplanets. Additionally, Brodholt et al (Brodholt et al., 2009) suggested that post-perovskite may have a much lower viscosity than perovksite, perhaps yielding a layered convection mantle for post-perovskite-mantle dominated planets. A layered convection would add additional thermal boundary layers raising the temperature of the core. Then, the large ambiguities concerning initial composition of exoplanets and physical properties of silicates at extreme conditions make thermal models (Sotin et al., 2007; Valencia et al., 2007; van den Berg et al., 2010) dependent of initial hypotheses. So conclusions concerning the state of exoplanets' cores are largely dependent on those thermal models. We believe that the dispersion shown in Fig. 3 reflects this degree of uncertainty within reasonable limits.

The thermal structure of super-Earths can then be compared with the melting temperatures of pure iron as calculated in this study (Figure 3). This does not account, however, for the effect of light elements alloying with iron. A list of light elements with high abundances and affinities with the metallic phase during planetary differentiation has been established: O, S, Si and C (Poirier, 1994). Eutectic melting occurring in binary systems could occur at temperatures lower by 1500 K than in pure iron (Fe-S at 65 GPa (Morard et al., 2008)), or only 200 K (Fe-Si at 21 GPa (Kuwayama and Hirose, 2004) and Fe-O at 50 GPa (Seagle et al., 2008)). On the basis of melting entropies calculated from *ab initio* calculations, a depression of the melting point of 700 K ± 100 K has been proposed (Alfè et al., 2002a). However, a notable discrepancy exists at the ICB conditions between pure iron melting temperature and the geotherm presented here (Figure 3 inset). Recent geotherm calculations (Hernlund and Labrosse, 2007) try to reconcile results on iron melting curve (Alfè et al., 2002b), implying a large temperature jump of ~2500K at the CMB. An exact discussion about the effect of different elements on the iron melting point would require more data at very high pressures. In the present study, a maximum melting point depression of 1500 K has been conservatively proposed (Figure 3).

Comparison of the thermal profiles with the melting curves taking into account all the aforementioned uncertainties reveals that the presence of liquid metal in the cores of terrestrial super-Earths is not likely for planets with masses superior to 2 $M_E$ (Valencia et al., 2007; van den Berg et al., 2010) (Figure 3). These values are also largely dependent on the melting depression due to light elements, taken here as 1500 K. In any case, the melting curve obtained in this study exhibits a steeper slope than planetary adiabatic temperature profiles. Thus, the more massive the planet, the more additional heat sources it would need to have a molten core.

Magnetic fields in terrestrial planets are classically based on a dynamo effect generated by rotation/convection in liquid metallic cores and would thus be unlikely in a massive super-Earth. The current study which provides the first calculation of the high pressure melting curve of iron at conditions encountered in the cores of terrestrial planets larger than Earth suggests that a magnetic field generated in a rotating-convecting liquid core may not be present in bodies a few times more massive than the Earth. The alternative of a "surface" magnetic field produced by water convection in a salty ocean, as it has been proposed for Ganymede (Schubert et al., 1996), must however be mentioned in the case of a large planetary ocean component (Valencia et al., 2007).

**Conclusion**

New constraints on the iron melting curve were obtained by AIMD calculations, using the two-phases method, highlighting a constant increase up to 14500 K ± 500 K at around 1500 GPa. Thermal structure of the Super-Earths, derived from previous publications (Sotin et al., 2007; Valencia et al., 2007; van den Berg et al., 2010), is discussed and compared to the calculated melting curve. A potential depression of the melting curve by 1500 K is also presented, in order to simulate light element effects on melting properties of iron. It appears that, for masses superior to 2 $M_E$, the Super Earths' cores are entirely solid; therefore no metallic-core generated magnetic field can be generated. The iron melting curve obtained in this study exhibits a steeper slope than any calculated planetary adiabatic temperature profile. Therefore, more additional heat sources will be required for maintaining molten metallic cores in massive terrestrial-type planets.

**Acknowledgments**

This work was founded by the SECHEL program of the *Agence Nationale de la Recherche* (Grant ANR-07-BLAN-185577)**.** The authors would like to thank the GENCI consortium for time provided on their supercomputers. The original manuscript was improved by the constructive comments of T. Spohn and three anonymous reviewers.

**Figure Caption**

Figure 1: High pressure melting curves of iron calculated using the two-phases approach (TPA) and heat until melt (HUM) approaches and compared with previous works at lower pressure (Alfè, 2009; Alfè et al., 2002b; Belonoshko et al., 2000). In these studies, iron structure investigated is hcp. Inset: zoom around the ICB pressures.

Figure 2: Melting curve of iron using Heat Until it Melts (HUM) method, using bcc (circle symbols) and hcp (square symbols) structures of iron. The melting temperatures depend on the numbers of atoms used in the simulation cell (54 (dashed lines) or 128 (solid lines) atoms), but no correlation is evidenced with the structure. Therefore, it is impossible in this study to assess the most stable structure of pure iron, as the effect of structure on the calculated melting point is small.

Figure 3: Comparison between the melting curves of iron obtained in this study and the thermal structure of terrestrial-type exoplanets (Sotin et al., 2007; Valencia et al., 2007; van den Berg et al., 2010). The uncertainty on the calculation of the inner core temperature is discussed in the text. The melting curve with a depression of 1500 K, representing the estimated maximum melting point depression, has been extended up to 2000 GPa. The inset is focused on the Earth's P-T conditions. The effect of different light elements is shown (5% for O and Si (Kuwayama and Hirose, 2004; Seagle et al., 2008); 30% for S (Morard et al., 2008)).

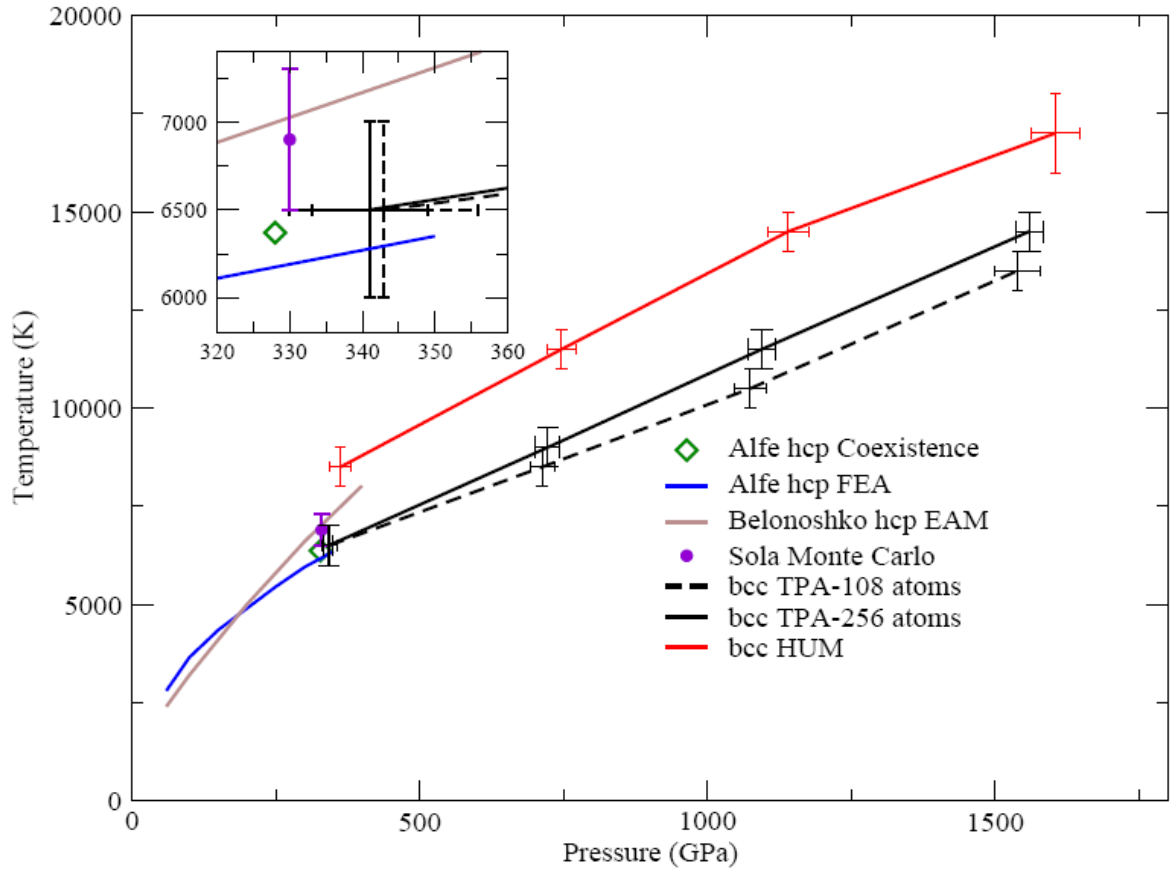

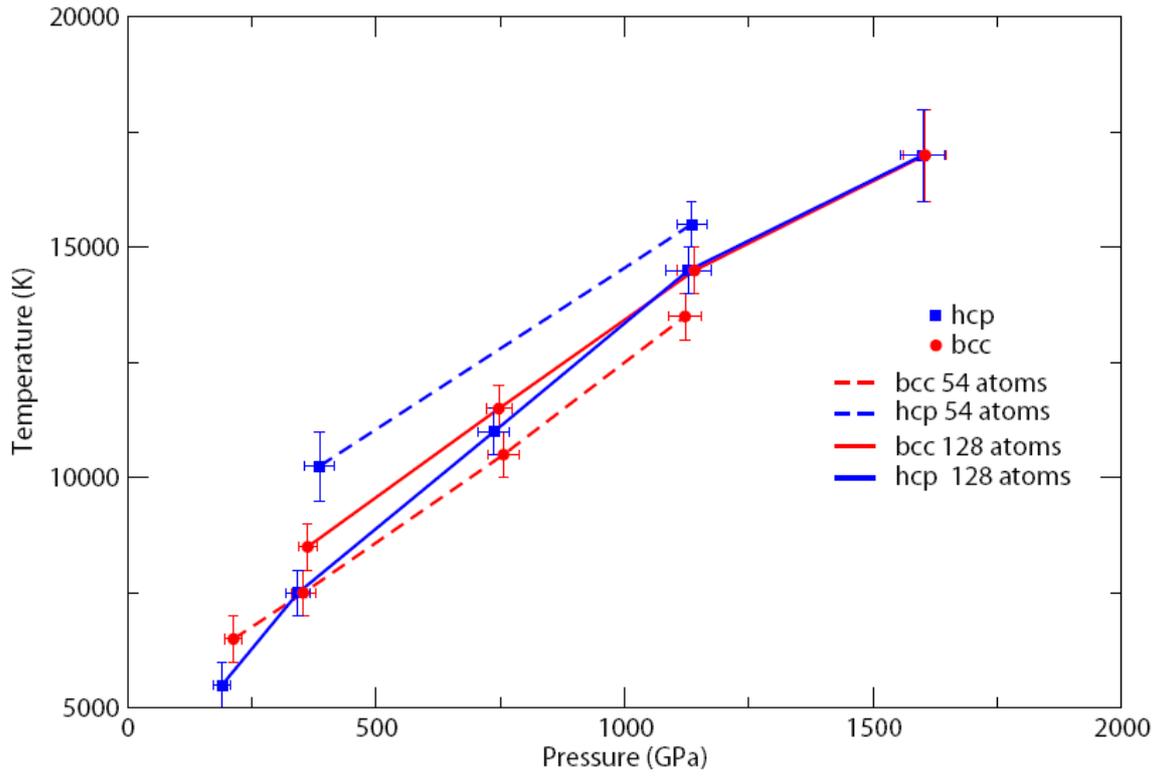

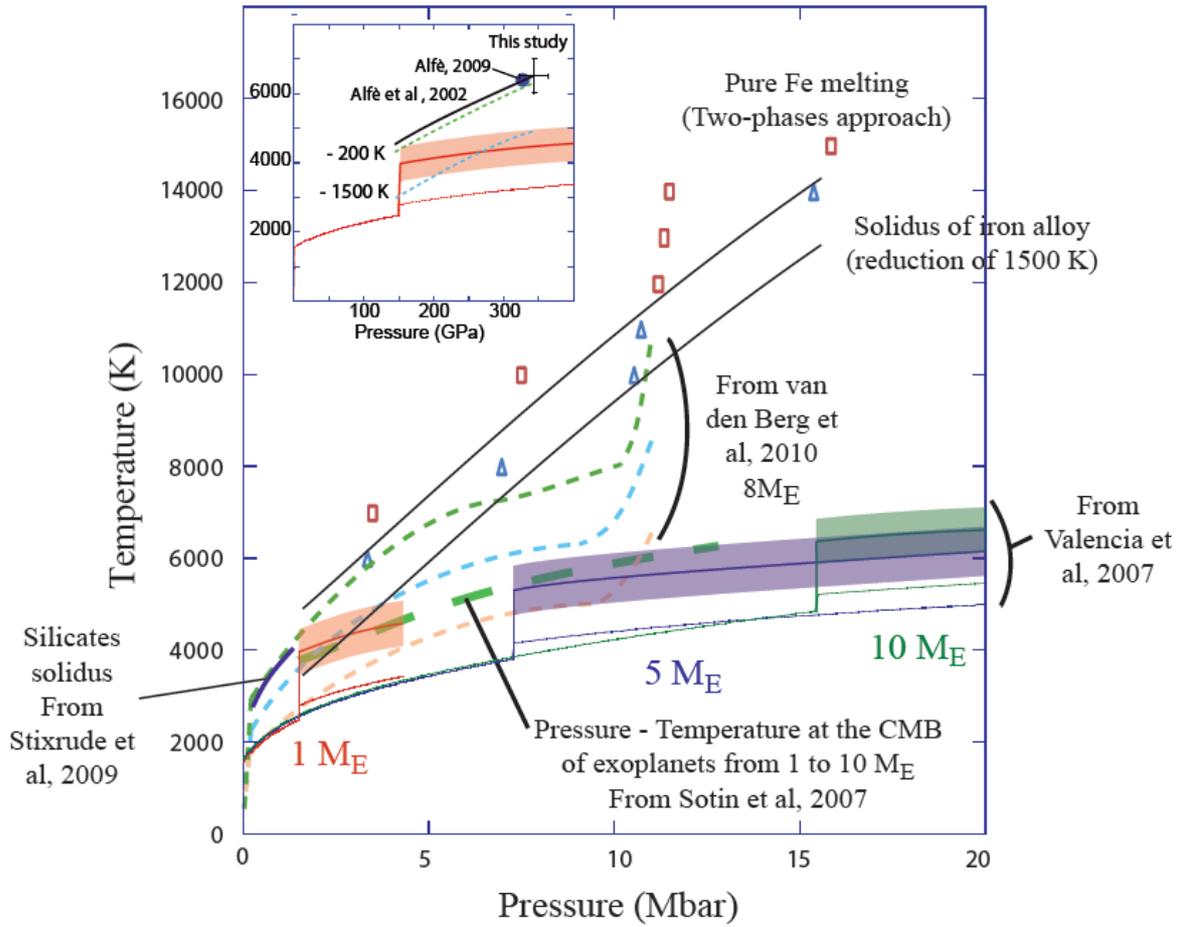